\documentclass[aps,pra,twocolumn,showpacs,amsmath]{revtex4-1}
\usepackage{graphicx}
\usepackage{amssymb}

\newcommand{\bra}[1]{\langle #1 | \,}
\newcommand{\ket}[1]{\, | #1 \rangle}

\newcommand{\ii}{\mathrm{i}}
\newcommand{\be}{\begin{equation}}
\newcommand{\ee}{\end{equation}}
\newcommand{\bea}{\begin{eqnarray}}
\newcommand{\eea}{\end{eqnarray}}
\newcommand{\besa}{\begin{subeqnarray}}
\newcommand{\eesa}{\end{subeqnarray}}
\newcommand{\bean}{\begin{eqnarray*}}
\newcommand{\eean}{\end{eqnarray*}}

\newcommand{\mbr}{\mathbf{r}}
\newcommand{\mbk}{\mathbf{k}}
\newcommand{\mbp}{\mathbf{p}}
\begin{document}

\title{Tan's distributions and Fermi-Huang pseudopotential in momentum space}

\author{Manuel Valiente}
\affiliation{Lundbeck Foundation Theoretical Center for Quantum System Research, Department of Physics and Astronomy, Aarhus University, DK-8000 Aarhus C, Denmark}

\date{\today}

\begin{abstract}
The long-standing question of finding the momentum representation for the s-wave zero-range interaction in three spatial dimensions is here solved. This is done by expressing a certain distribution, introduced in a formal way by S. Tan [Ann. Phys. {\bf 323}, 2952 (2008)], explicitly. The resulting form of the Fourier-transformed pseudopotential remains very simple. Operator forms for the so-called Tan's selectors which, together with Fermi-Huang pseudopotential, largely simplify the derivation of Tan's universal relations for the Fermi gas, are here derived and are also very simple. A momentum cut-off version of the pseudopotential is also provided, and with this no apparent contradiction with the notion of integrals in Tan's methods is left. The equivalence, even at the intermediate step level, between the pseudopotential approach and momentum-space renormalization of the bare Dirac delta interaction is then shown by using the explicit form of the cut-off pseudopotential.

\end{abstract}


\maketitle

\section{Introduction}
The physics of strongly interacting quantum many-body systems has been one of the most active fields of research for a number of decades now \cite{FetterWalecka}. These systems, which are relevant in many areas of physics, such as atomic, molecular and optical \cite{BlochReview}, condensed matter \cite{Bruus} or nuclear physics \cite{RingSchuck}, offer challenging theoretical and experimental problems that are attracting much of recent interest.

Lately, one of the most studied problems is that of a spin-$1/2$ many-fermion system interacting with short-range potentials with a large two-body scattering length, i.e. close to unitarity. When this is the case, perturbation theory is not useful due to the absence of an obvious small parameter in the system, and solving the problem becomes a major challenge. In this direction, quantum Monte Carlo and density-functionals \cite{Bulgac1}, as well as the epsilon expansion \cite{Nishida} have been succesfully applied. In addition, some of the most striking properties of interacting Fermi gases at unitarity are those related to its universality \cite{Ho}, due to the absence of a natural length scale associated with the interaction. 

Close to but away from unitarity, the so-called Tan's relations \cite{Tanenergetics,Tanmomentum,Tanvirial} dictate the behavior of the many-fermion system. These important results relate many of the many-body properties, such as the adiabatic change of the energy when varying the two-body scattering length, the asymptotic momentum distribution and the pressure to a single quantity called the contact, and were recently verified experimentally by Stewart {\it et al.} \cite{Stewart}. In his pioneering works, S. Tan introduced, quite formally, the regularized s-wave zero-range, or Fermi-Huang \cite{Fermi,Huang} pseudopotential in momentum representation. He also introduced a set of distributions -- selectors -- in the same fashion of his approach to the interaction. These distributions proved very useful in the derivation of Tan's relations, radically simpler than with conventional approaches used later, such as the operator product expansion \cite{OPE}. The tools developed in \cite{Tanenergetics} seem, at first sight, to run into contradiction, for a new notion of improper integrals or the existence of new types of generalized functions -- distributions -- was needed in order to introduce them.

In this Letter, by explicitly constructing Tan's generalized functions, we find that they are nothing but usual distributions containing Dirac deltas. Moreover, related distributions -- which become the desired generalized functions after taking limits -- are constructed so that integrals over the whole momentum space can be defined as limits of cut-off integrals, as in ``usual'' mathematics. These findings put an end to any kind of controversy or contradiction regarding Tan's approach \cite{Braaten}. Moreover, they solve the long-standing problem (54 years old!) of finding the Fourier-transformed Fermi-Huang pseudopotential, which may be useful for few- and many-body problems worked out in momentum representation, and for variational or perturbative calculations done in this way. As an important corollary of these results, it is proven that renormalization of the Dirac delta interaction, and Fermi-Huang pseudopotential in momentum space are equivalent to one another, even at intermediate steps in calculations.

\section{Fermi-Huang pseudopotential}
In order to have a model potential which is simple enough, still being able to reproduce the low-energy scattering properties in a system, the regularized zero-range s-wave (Fermi-Huang) pseudopotential \cite{Huang} is useful. This two-body interaction has the form
\be
V(\mbr-\mbr')= g \delta(\mbr-\mbr')\frac{\partial}{\partial |\mbr-\mbr'|}(|\mbr-\mbr'|\cdot).\label{pseudo}
\ee
The differential term multiplying the Dirac delta has the effect of removing $1/|\mbr-\mbr'|$-type singularities at the contact between two particles. Actually, the derivative in (\ref{pseudo}) suggests that if it was absorbed into the definition of the coupling constant $g$, then this would depend on the relative momentum $k$ (operator \cite{ValienteHard}), probably in a very special way. This will clearly be one way of interpreting the results found in the following section.

\section{Pseudopotential in momentum space} In Tan's pioneering work \cite{Tanenergetics}, the formal momentum representation of the s-wave Fermi-Huang pseudopotential $\delta(\mathbf{r})\partial_r(r\cdot)$ is obtained by defining the $\Lambda$-distribution via
\be
\delta(\mbr)\frac{\partial}{\partial r}(r e^{\ii \mbk \cdot \mbr})\equiv \delta(\mbr)\Lambda(\mbk),\label{Lambda1}
\ee
where $\Lambda$ is defined as $\Lambda(\mbk)=1$ for $k<\infty$ together with the following integral relation
\be
\int \mathrm{d}\mbk k^{-2}\Lambda(\mbk) = 0. \label{Lambda2}
\ee 
A na\"ive, rather illuminating interpretation of the above equations is that $\Lambda$ is equal to 1 everywhere except at infinity, where it behaves as a distribution and not as a usual function. Moreover, relation (\ref{Lambda2}) has a clear physical meaning: the two-body scattering length $a$ diverges at infinitely large coupling constant $g$; indeed, as is well known, $g\propto a$.

The choice for $\Lambda$ that can be made to satisfy both Eqs. (\ref{Lambda1}) and (\ref{Lambda2}) has the form
$\Lambda(\mbk)= 1 - \mathcal{G}(\mbk)\delta(\mathcal{F}(\mbk))$, where $\mathcal{G}$ and $\mathcal{F}$ are ordinary real functions of the momentum, and have a singularity at $\mbk = 0$. Introducing this choice into (\ref{Lambda2}), we obtain $\mathcal{G}(\mbk)=\mathcal{F}(\mbk)=1/k$, where the integral in Eq. (\ref{Lambda2}) is understood after performing the change of variables $1/k=\tau$. The explicit form of the $\Lambda$-distribution is therefore given by
\be
\Lambda(\mbk)=\Lambda(k)=1-\frac{\delta(1/k)}{k}.\label{Lambdafin}
\ee
This is the main result of this Letter. In particular, it means that the momentum-space representation of the two-body pseudopotential has the form $V(\mbk,\mbk')/g=1-\delta(1/|\mbk'|)/|\mbk'|$; $V$ is obviously non-Hermitian -- and so it is with $\Lambda$ as defined by Tan \cite{Tanenergetics}, Eq. (\ref{Lambda2}) -- but this represents no problem and is analogous to what happens to the one-dimensional hard-sphere Bose gas recently discussed in \cite{ValienteHard}.  Eq. (\ref{Lambdafin}) puts an end to the long-standing quest of finding the explicit form of the Fermi-Huang pseudopotential in momentum space.
It is now clear what $\Lambda$ actually does. First of all, it lets the interaction be the bare zero-range two-body potential at finite momenta, while the singular, momentum-space delta-like behavior removes the ultraviolet divergence of integral (\ref{Lambda2}) at infinity. 

Tan also introduced a set of useful distributions \cite{Tanenergetics}, which he called selectors $L(\mbk)$ and $\eta(\mbk)$, that fundamentally simplify the derivation of his celebrated relations \cite{Tanenergetics,Tanmomentum,Tanvirial}. The $L$-selector is defined as zero for any $k<\infty$, but satisfies
\be
\int \frac{\mathrm{d}\mbk}{(2\pi)^3}\frac{L(\mbk)}{k^2} =1.
\ee 
Following a similar analysis as for the $\Lambda$-distribution, we obtain the explicit form of the $L$-selector,
\be
L(\mbk)=2\pi^2\frac{\delta(1/k)}{k^2},\label{Lselector}
\ee
and all its properties, derived by Tan in \cite{Tanenergetics}, follow easily from the above relation. In addition, its position representation $\ell (\mathbf{r})$ is given by
\be
\ell(\mathbf{r})\equiv\int \frac{\mathrm{d}\mbr}{(2\pi)^3} L(\mbk)e^{i\mbk\cdot \mathbf{r}} = -4\pi \delta(\mbr)r^2\frac{\partial}{\partial r}.
\ee
The $\eta$-selector, defined as $\eta(\mbk)=\Lambda(\mbk)+L(\mbk)/2\mu g$, with $\mu$ the two-body reduced mass, has therefore the following form
\be
\eta(\mbk)=1-\frac{1}{\mu g k^2}(\mu gk - \pi^2)\delta(1/k).
\ee 

We are now in position of writing down, for instance, Tan's energy theorem \cite{Tanenergetics} for a (homogeneous) spin-$1/2$ Fermi gas with contact interactions between particles of different spin, totally explicitly as
\be
E=\hbar^2\sum_{\mbk,\sigma}\left[\frac{k^2}{2m}-\left(\frac{k}{2m}-\frac{\pi^2}{m^2 g}\right)\delta(1/k)\right]n_{\mbk,\sigma},
\ee
with $E$ the energy of the system, the sum in $\sigma$ running through the two possible spin components $\uparrow$,$\downarrow$, and $m=2\mu$ the single-particle mass. The Dirac delta in the sum has to be regarded, of course, as a limit representation before the thermodynamic limit is taken where the sum becomes an integral.

\section{Cut-off momenta and well-defined limits of improper integrals}
In practice, e.g. in numerical calculations, a momentum cut-off $k_c<\infty$ is often required. However, the pseudopotential in momentum space needs the evaluation of integrals over infinite space, as seen in Eq. (\ref{Lambda2}). This fact triggered the claim \cite{Tanenergetics} that a new notion of integrals was needed to define Tan's distributions. This problem is already absent if we use Eqs. (\ref{Lambdafin}) and (\ref{Lselector}) but, in order to make finite integrals with the pseudopotential have a sensible meaning, we need to define a cut-off $\Lambda$-distribution in the following way:
\be
\Lambda^{k_c}(k)=\theta(k_c-k)-\frac{\delta(1/k-1/k_c)}{k},\label{Lambdacut}
\ee
where $\theta(k)$ is the Heaviside step function. Obviously, $\Lambda=\lim_{k_c\to \infty}\Lambda^{k_c}$. It is easy to see that $\Lambda^{k_c}=1$ for $k<k_c$ and
\be
\int \mathrm{d}\mbk k^{-2}\Lambda^{k_c}(\mbk)=\int_{|\mbk|\le k_c} \mathrm{d}\mbk k^{-2}\Lambda^{k_c}(\mbk) = 0,\label{Lambda3}
\ee
which is the cut-off version of (\ref{Lambda2}). Similar cut-off versions for the $L$ and $\eta$ selectors are trivially obtained in the same way. From Eq. (\ref{Lambda3}), we see that the ``problem'' of infinite integrals not being limits of finite integrals \cite{Tanenergetics} is non-existent if we use $\Lambda^{k_c}$.
Intuitively, this cut-off-dependent interaction brings down the infinity to the value of the cut-off chosen for a particular computation. It may also be replaced by a convenient limit representation with usual functions in numerical calculations.

\section{Equivalence with momentum-space renormalization}
We proceed now to see how the above procedure is totally equivalent to the usual renormalization scheme for the delta interaction. Rather complete treatments of renormalization with effective interactions are given in \cite{Kolck,Lee,Kaplan}. Consider the stationary Schr\"odinger equation $H\psi=E\psi$ in the relative coordinate, with
\be
H=\frac{\mbp^2}{2\mu}+g\delta(\mbr)\frac{\partial}{\partial r}(r\cdot).\label{TwoBodyHam}
\ee
We consider bound-state ($E<0$) solutions in the momentum representation, and we set $\hbar\equiv 1$ throughout. The integral equation for the energy $E=-|E|$ has the form
\be
1=-\frac{g}{2\pi^2}\int_{0}^{\infty}\mathrm{d}k \frac{k^2\Lambda(k)}{k^2/2\mu + |E|}.\label{inteq}
\ee
In order to evaluate the above integral without making use of the formal properties of $\Lambda$ \cite{Tanenergetics}, we note that, using Eq. (\ref{Lambdacut}), 
\begin{align}
&\int_{0}^{\infty}\mathrm{d}k \frac{k^2 \Lambda(k)}{k^2/2\mu + |E|}=\label{green}\\
&\lim_{k_c\to \infty}\left[ \int_{0}^{k_c} \mathrm{d}k \frac{k^2}{k^2/2\mu+|E|} - \int_{1/k_c}^{\infty} \mathrm{d}\tau \frac{\delta(\tau-1/k_c)\tau^{-1}}{(2\mu)^{-1}+|E|\tau^2}\right],\nonumber
\end{align}
where we have changed variables as $k=1/\tau$ in the second integral. From Eqs. (\ref{inteq}) and (\ref{green}) we obtain
\begin{align}
&\lim_{k_c\to \infty} \left[\sqrt{2\mu|E|}\arctan\left(\frac{k_c}{\sqrt{2\mu |E|}}\right)
-k_c\left(1-\frac{1}{1+\frac{2\mu |E|}{k_c^2}}\right)\right]\nonumber \\
&=\frac{\pi^2}{\mu g},\label{nonren}
\end{align}
which gives, after using $g=2\pi a/\mu$, the bound state energy $E=-1/2\mu a^2$, as should be.

The first of the integrals in the second line of Eq. (\ref{green}) is the one appearing in the ill-defined problem of finding the bound state for the (irregular) delta interaction $g_B\delta(\mbr)$,
\begin{align}
-\frac{\pi^2}{\mu g_B}&=\lim_{k_c\to \infty} \int_{0}^{k_c} \mathrm{d}k \frac{k^2}{k^2/2\mu+|E|}\label{ren}\\
&=\lim_{k_c\to \infty} \left[k_c-\sqrt{2\mu|E|}\arctan\left(\frac{k_c}{\sqrt{2\mu|E|}}\right)\right].\nonumber
\end{align}
Here, the interaction strength $g_B$ plays the role of the so-called bare coupling constant. The renormalized coupling constant $g_R$ is then defined as 
\be
\frac{1}{\mu g_R}\equiv \frac{1}{\mu g_B} +\frac{k_c}{\pi^2}, \label{rencon}
\ee
and the limit $k_c\to \infty$ is taken while keeping $g_R$ constant. We obtain a cancelation of the linearly divergent term $k_c$, and finally
\be
E=-\frac{2\pi^2}{\mu^3 g_R^2}.\label{EnergygR}
\ee
It is now obvious that if the renormalized coupling constant $g_R$ coincides with the original Fermi-Huang coupling constant $g$, then Eq. (\ref{EnergygR}) becomes $E=-1/2\mu a^2$, and coincides with the energy calculated with the $\Lambda$-distribution. Actually, from Eqs. (\ref{nonren}-\ref{rencon}), we observe that the use of Fermi-Huang pseudopotential in momentum space and the renormalization procedure defined in (\ref{rencon}) are equivalent -- except for a negligible factor of $(1+2\mu|E|/k_c^2)^{-1}$ which we could arbitrarily include in the renormalization -- provided that $g_R\equiv g$. This last fact is of course natural, since the bound state energy $E=-1/2\mu a^2$ is the fit parameter used in renormalization approaches; actually, both renormalization and Fermi-Huang pseudopotentials aim at fitting the bound-state energy (or equivalently the scattering length).

\section{Perturbation theory} We show now how perturbation theory can be implemented using the $\Lambda$-distribution without the need of perturbative renormalization, and that such two approaches are again equivalent to one another.

Let us consider two-body scattering from the Fermi-Huang pseudopotential, described in position representation by Hamiltonian (\ref{TwoBodyHam}) in the relative coordinate. The system's t-matrix $T(z)$, $z\in \mathbb{C}$, is the solution to the Lippmann-Schwinger equation $T(z)=V+VG^{(0)}T(z)$, with $G^{(0)}\equiv G^{(0)}(z)$ the non-interacting Green's function. In momentum space, the matrix elements $\bra{\mathbf{p}'}T(z)\ket{\mathbf{p}}=T(z;\mathbf{p}',\mathbf{p})$ we calculate are at finite momenta $\mathbf{p}'$ and $\mathbf{p}$. Therefore the only elements that require the regularizing term $\delta(1/k)/k$ are those which involve $k\to \infty$, that is, only $\bra{\mathbf{p}'}V\ket{\mathbf{k}}=V(\mathbf{p}',\mathbf{k})$ when it is in an integral over $\text{d}\mathbf{k}$. Taking this fact into account, we can calculate the perturbative expansion (Born series) for the t-matrix, with all its terms being finite. Note that due to the special form of the contact interaction, $T(z;\mathbf{p}',\mathbf{p})=T(z)$ is independent of $\mathbf{p}'$, $\mathbf{p}$.

On-shell ($z=q^2/2\mu+i\eta$), the first order term $t^{(1)}$ of the Born series is given by $t^{(1)}=V(\mathbf{p}',\mathbf{p})=g$. It is identical to the first Born obtained for a bare delta interaction after first-order renormalization $g_B \to g_R = g$ (which involves no infinities). The second order term $t^{(2)}$ is easily calculated as 
\be
t^{(2)}(q^2/2\mu+i\eta)=\frac{g^2}{2\pi^2} \int_{0}^{\infty}dk k^2 \frac{1-k^{-1}\delta(1/k)}{(q^2-k^2)/2\mu + i\eta}.
\ee
The above quantity is found to be finite, and is given by $i\mu g^2 q/2\pi^2$. The reader can easily check that second-order perturbative renormalization for a bare Dirac delta interaction yields exactly the same result, provided that the renormalized coupling constant $g_R=g$. We may continue the perturbative expansion to arbitrary order to find that all terms $t^{(n)}(q^2/2\mu+i\eta)=(i\mu gq/2\pi)^{n-1} g$ coincide with those obtained through perturbative renormalization, and resum the resulting series to obtain the exact t-matrix $T(q^2/2\mu+i\eta)/g=(1 - i \mu g q/2\pi^2)^{-1}$.

There is yet another way of calculating the t-matrix perturbatively. First note, from its exact form, that it is holomorphic in the inverse coupling constant $\lambda\equiv (2\pi)^3/g$. We can therefore expand it in powers of $\lambda$. We may also obtain the series by calculating exactly only the t-matrix at infinite coupling constant $T_{\infty}(z)$. This may be as well considered as a different renormalization prescription, where the on-shell $T_{\infty}$ is measured in an experiment, although regularization-renormalization of infinities is here not required. Since $\Lambda=V/g$, we have $\Lambda G^{0}T_{\infty}(z)=-1$, which is solved on-shell by $T_{\infty}(q^2/2\mu + i\eta)=2\pi^2 i/\mu q$; formally, $T_{\infty}(z)=-(\Lambda G^{(0)})^{-1}$.  We immediately obtain $T=T_{\infty} - \lambda T_{\infty} T$, which is easily solved perturbatively by $T=T_{\infty}\sum_{n}(-\lambda T_{\infty})^n$. After resummation the t-matrix is given by $T=T_{\infty}/(1+\lambda T_{\infty})$, the exact result also obtained above.

\section{Concluding remarks}
We have obtained the explicit momentum-space representation of the zero-range s-wave pseudopotential and Tan's selectors. Their forms are very simple and involve usual Dirac delta distributions only. This implies that the generalized functions defined by Tan in \cite{Tanenergetics} do not constitute a novel kind of distributions, therefore clearly justifying the methods he used to derive his celebrated relations \cite{Tanenergetics,Tanmomentum,Tanvirial} for the Fermi gas. We have also constructed a cut-off version of the pseudopotential, which may be useful for finite-size computations in momentum space. The cut-off pseudopotential is crucial to make improper integrals involving Tan's distributions become limits of integrals over a finite interval, further justifying Tan's methods. We have then shown that the use of this potential or a renormalization procedure are totally equivalent, both approaches aiming at removing ultraviolet divergences, and we have exemplified this with the two-body bound state problem and perturbation theory.

Our results constitute a promising path towards obtaining the momentum representation of higher partial wave pseudopotentials \cite{Calarco}, which can be of great interest for the many-body problem in momentum space with such interactions; this could simplify many calculations, as in the presently studied case of s-wave interaction. 
An important implication of the results of this Letter is that the derivation of many celebrated results in the literature may become more accessible for a non-specialized audience having only a working knowledge of or a preference for explicit operator methods.

It would be very interesting to investigate the explicit form of three-body contact interactions for bosons in momentum space leading to the equivalents of different three-body renormalization schemes \cite{PricoupenkoReno,3BodyReno}.

\acknowledgments
I wish to thank Klaus M{\o}lmer for encouragement and support. The author acknowledges financial support from a Villum Kann Rasmussen block scholarship.

\end{document}